  \documentclass{agujournal2018}
 \usepackage{apacite}
\usepackage{url} %this package should fix any errors with URLs in refs.
 \draftfalse
\journalname{Earth and Space Science}
\begin{document}
\title{Helicity dynamics, inverse  and bi-directional cascades in fluid and magnetohydrodynamic turbulence: \\
A brief review}

\authors{A. Pouquet\affil{1,2}, D. Rosenberg\affil{3}, J.E. Stawarz\affil{4} \ and R.  Marino\affil{5}} 
%\thanks{Also funded by NASA, NOAA, NSF, ....}}
\affiliation{1}{Laboratory for Atmospheric and Space Physics, Boulder, CO 80309 USA.}
\affiliation{2}{NCAR, P.O. Box 3000, Boulder, CO 80307 USA.}
\affiliation{3}{NOAA, Boulder, CO, United States.}
\affiliation{4}{Department of Physics, Imperial College London, United Kingdom.}
\affiliation{5}{Laboratoire de M\'ecanique des Fluides et d'Acoustique, CNRS, \'Ecole Centrale de Lyon,
  Universit\'e de Lyon, INSA de Lyon-1, F-69134 \'Ecully, France}
\correspondingauthor{A. Pouquet}{pouquet@ucar.edu}

\begin{keypoints} % 140 characters max each 
\item % 139 char
Magnetic helicity displays an inverse cascade to large scales which, in electron MHD, can be justified with a simple phenomenological model  
\item % 140 char
Total energy in MHD or rotating stratified turbulence has constant-flux cascades to small and large scales, affecting mixing and dissipation  
\item % 140 char
With these results, needed modifications to sub-grid scale turbulence models will be enhanced using tools from big-data and machine-learning  
\item
\item
\item Published on line, {\it Earth and Space Science}, AGU/Wiley, March 6, 2019. 
\item Special Section: Nonlinear Systems in Geophysics: Past Accomplishments and Future Challenges
\item DOI:10.1029/2018EA000432.

% Nonlinear dynamics, ubiquitous in both physical and social sciences, leads to extreme events whose complexity engenders strong perturbations to eco-systems. Assessing and predicting earthquakes and floods, tornadoes and hurricanes, weather and climate, or modeling anomalous dissipation, mixing and bursty reconnection are performed using a variety of tools. Diagnostic and prognostic tools include stochastic equations, laboratory experiments, observations, numerical simulations and analysis, statistics and big data retrievals, as well as mathematical and phenomenological modeling. Fractals, and scaling laws in general, are one such tool which will be reviewed in depth, with applications to the interior of the Earth, hydrology, the ocean, the atmosphere and heliophysics. These tools will continue to evolve: How? And where will we be going with them? What are the challenges? In the framework of AGUÕs Centennial, this Special Issue- to which we invite contributions - will assess where we are and what are the perspectives for progress.

% Special section organizers: Cecile Penland, NOAA/Earth System Research Lab; Leonard Pietrafesa, North Carolina State University; Annick Pouquet, Computational and Information Systems Laboratory, NCAR, and Laboratory for Atmospheric and Space Physics, University of Colorado at Boulder; Sarah Tebbens, Department of Physics, Wright State University
\item

%                                                                                                               \item {\bf  SUBMIT: Copy bbl file}!
% \hskip0.0777truein Annick: http://orcid.org/0000-0002-2355-2671

\end{keypoints}

\begin{abstract}     % 250 words max, 250 here 
We review helicity dynamics, inverse and bi-directional cascades in fluid and magnetohydrodynamic (MHD) turbulence, with an emphasis on the latter. The energy of a turbulent system, an invariant in the non-dissipative case, is transferred to small scales through nonlinear mode coupling. Fifty years ago, it was realized that, for a two-dimensional fluid, energy cascades instead to larger scales, and so does magnetic excitation in MHD. However, evidence obtained recently indicates that in fact, for a range of governing parameters, there are systems for which their ideal invariants can be transferred, with constant fluxes, to both the large scales and the small scales,
as for MHD or rotating stratified flows, in the latter case including with quasi-geostrophic forcing. 
Such bi-directional, split, cascades directly affect the rate at which mixing and dissipation occur in these flows in which nonlinear eddies interact with fast waves with anisotropic dispersion laws, due for example to imposed rotation, stratification or uniform magnetic fields. The directions of cascades can be obtained in some cases through the use of phenomenological arguments, one of which we derive here following classical lines in the case of the inverse magnetic helicity cascade in electron MHD. 
With more highly-resolved data sets stemming from large laboratory experiments, high-performance computing  and {\it in-situ} satellite observations, machine-learning tools are bringing novel perspectives to turbulence research. Such algorithms help devise new explicit sub-grid scale parameterizations, which in turn may lead to enhanced physical insight, including in the future in the case of these new bi-directional cascades. 
\end{abstract}

\noindent {\bf Plain-language Summary}  % 200 words max, 195 here 
Turbulent flows are ubiquitous in Geophysics and Space Physics. They are complex, involving interactions between eddies and waves at widely separated scales, with the energy flowing in the general case only to small scales to be dissipated. It was found recently that, contrary to such expectations, energy can go in substantial amounts to both the small scales and to the large scales, in the presence of magnetic fields, as applicable to space plasmas as in the Solar Wind, and for rotating stratified flows as encountered in the atmosphere and the oceans. This result implies that the amount of energy available for dissipation may differ from flow to flow and simple scaling arguments allow for predictions that are backed up by results stemming from direct numerical simulations. One should incorporate this bi-directional cascade phenomenon in the turbulence models used for global computations of geophysical and astrophysical media. In fact, machine-learning tools may prove useful in deriving such enhanced models in their capacity to interrogate the large numerical, observational and experimental data bases that already exist for such complex flows, with the potential to lead to a deeper understanding and to more accurate predictions of such flows.

\section{Introduction} \label{S:intro}  

Turbulence prevails in many geophysical and astrophysical flows, and progress is being made presently in the understanding of such flows for fluids, neutral or conducting, including in the presence of waves stemming from strong rotation, stratification, compressibility or quasi-uniform magnetic fields. The complexity of turbulent flows comes from the nonlinearity of the underlying equations leading to multi-scale interactions through mode coupling. It can result in a non-universality of spectra in models of magnetohydrodynamic (MHD) turbulence
 \citep{mininni_07b,  lee_10, beresnyak_14, perez_14}, 
 with applications to Solar Wind dynamics  \citep{galtier_12} or laboratory plasmas \citep{ bratanov_13}. It is also at the source of a wide variety of behavior, from the appearance at small scales of sharp structures such as tornadoes, to large-scale pattern formation, coherent vortices and jets. This type of ordered motion at large scale is not limited to physical systems. For example, 
 large-scale structures are
 also encountered in micro-biology in the context of the collective behavior of  constituents within so-called active fluids (see \citet{reinken_18} and references therein, in a fast-moving domain of research), involving interactions between the solvent (say, water) and self-propelled micro-organisms such as bacteria 
 %am\oe b\ae\  
\citep{dombrowski_04, wensink_12}. By collective behavior, it is meant the dynamics of a group viewed as an entity, a phenomenon that can be encountered in statistical physics, for example for the Ising model, as well as in other fields such as voter or crowd dynamics  \citep{castellano_09}.

Substantial advances in the understanding of nonlinear systems have dealt with the zero-dimensional case in which temporal chaos in an otherwise spatially ordered field is observed in many instances, together with  fractal behavior, leading to remarkable scaling laws (see several reviews in this Special Issue). Similarly, in one spatial dimension, solitons can arise through an exact balance between dispersion and  nonlinear steepening, due for example to advection, as in the Korteweg-de-Vries or Kuramoto-Sivashinsky equations. Finally,  in higher spatial dimensions, the seminal discovery, first for two-dimensional neutral fluids \citep{kraichnan_67} of the possibility of an inverse cascade has led to a fundamentally different view of turbulent flows beyond the venue that mode-coupling offers for energy dissipation in the  small scales. Inverse cascades are defined here as an excitation reaching scales larger than the forcing scale because of constraints due to the presence of more than one ideal quadratic invariant in the non-dissipative case.

Inverse and direct cascades are found in other systems, such as nonlinear optics. For example in \citet{newell_08}, a direct connection to dissipation at small scales through instabilities of coherent structures at large scales is stressed in such a context. Indeed, in the presence of an inverse cascade leading to instabilities of these large-scale coherent structures, one prediction of weak turbulence -- that is, turbulence with waves that are faster than nonlinear eddies -- is that there is stronger intermittency in the small scales. This enhanced intermittency can 
be linked to the presence of strong and sporadic small-scale structures such as vortex and density filaments in the interstellar medium, and current sheets in the Earth's magnetosphere.
It should be noted that the corrections to simple (dimensional) scaling laws for turbulent fluids can be computed analytically in some cases, such as for the passive scalar \citep{kraichnan_94}, and progress is being made using non-perturbative renormalization group theory for fully developed turbulence (FDT) \citep{canet_16}.
% take the form of wave breaking which is often observed for example at the surface of the ocean.
Moreover, in weak MHD turbulence in the presence of a strong uniform magnetic field, small-scale intermittency is found as well, together with nonlocal coupling of scales \citep{meyrand_15}. In weak Langmuir turbulence, such a coupling between large-scale coherent structures, formed by an inverse cascade in the form of Langmuir cavitons, and strong small-scale turbulence, is also advocated in the coupling of these two forms of turbulence \citep{henri_11}. Similarly, the non-Gaussian wings observed for example in four-wave interactions for Langmuir waves are associated with wave breaking, and intermittency is stronger when the linear and nonlinear characteristic  times become comparable \citep{choi_05b}. A  simple case of such a strong intermittency, when linear and nonlinear characteristic times are comparable, is that of the vertical velocity of rotating stratified flows; a model for it can be developed for a Philipps saturation spectrum 
 \citep{dewan_97}, as shown for example in recent direct numerical simulations (DNS)
  \citep{rorai_14, feraco_18}, or when the shorter time-scale of the problem is that associated with random sweeping \citep{clark_15b}. 

These new results are due to a combination of technological, observational, numerical and theoretical advances, with the help of several large-scale laboratory experiments such as the Coriolis table to study rapidly rotating turbulence, in the presence or absence of stratification \citep{aubourg_17}. Similarly, high values of the control parameter, namely the Reynolds number which measures the relative strength of nonlinearities to (linear) dissipation,  are achieved nowadays using liquid helium \citep{saint_14}. At the same time, a multitude of observations are increasing our understanding of such complex fluids, with various space missions \citep{marino_08, marino_11, tsurutani_16}, now including using data stemming from MMS (Magnetospheric Multi-Scale,  see {\it e.g.} \citet{burch_16}), looking at turbulence in the magnetotail \citep{ergun_18} or in the magnetosheath \citep{gershman_18}, as well as in the Solar Wind in general \citep{chasapis_17}.

Such studies of the Earth's plasma  environment can lead to advance warnings of strong solar eruptions and their ensuing disruption on satellite communications \citep{cassak_17}. In parallel, progress in high-performance computing is allowing for numerical simulations at Reynolds numbers that were not explored before with sufficient accuracy  in the absence of modeling terms \citep{debruynkops_15, rosenberg_15, ishihara_16, iyer_17, zhai_18}. 
And a renewal of interest in the theory of weak turbulence has allowed for the exploration of small-scale and large-scale intermittency as diagnosed in the occurrence of extreme events,  such as  fronts and filaments in the ocean \citep{mcwilliams_16}, or current sheets in MHD \citep{matt_rev_04, mininni_06b},  
as well as shocks in compressible flows \citep{porter_02}, all events making such flows largely unpredictable.
A picture of turbulence  is thus emerging that broadens our understanding which, in some instances, dates back to the mid forties for homogeneous isotropic  turbulence, for example concerning the possible scaling laws for energy spectra, classically compared to the Kolmogorov (1941) law for the kinetic energy spectrum, namely $E_V(k)\sim \epsilon_V^{2/3}k^{-5/3}$, where $k$ is the isotropic wavenumber and $\epsilon_V$ the kinetic energy dissipation rate. 
It is in this context that we now review results on helicity, and on inverse and bi-directional cascades in MHD and fluid  turbulence, with an emphasis on the former. 

\section{The role of kinetic helicity}   

We first recall  the general properties of kinetic helicity, $H_V$, in fluid turbulence. It is defined as the space-integrated correlation between the velocity field ${\bf u}$ and the vorticity ${\mbox{\boldmath $\omega$}} =\nabla \times {\bf u}$, and is an invariant of the ideal equations of motion for FDT, as well as for purely rotating flows. Helicity is not definite positive: it can be of either sign, representing a (partial) alignment or anti-alignment of  velocity and vorticity. Helicity has been studied extensively in the laboratory and in numerical simulations  \citep{moffatt_92, biferale_12}; it is strong in vortex filaments in FDT \citep{kerr_85, kerr_87}, as well as in quantum turbulence \citep{clark_17}. It has recently been detected in DNS using helicoid particles
 \citep{gustavsson_16}, particles which are sensitive to the lack of mirror symmetry of the flow in which they are embedded. 
The decay of energy is considerably slowed down in the presence of strong helicity  for either rotating  turbulence \citep{teitelbaum_11}, or in the stratified case \citep{rorai_13}. The helicity spectrum has been observed in the Planetary Boundary Layer \citep{koprov_05}; it is found to be rather flat when the stable stratification is strong, a finding also present in DNS \citep{rorai_13}. Helical modes have also been identified in secondary instabilities in boundary layer flows leading to the formation of turbulent spots \citep{bose_16}.
 
In Hurricane Bonnie (1998), velocities, helicity and brightness temperature were monitored with tropospheric drop sondes  \citep{molinari_08, molinari_10}. Hurricanes are associated with strong helicity since, outside the boundary layer, they can be viewed as a quasi two-dimensional rotating stratified flow, thus with mostly vertical vorticity, together with a strong updraft or downdraft. 
 Apart from the atmosphere, helicity in geophysical flows is observed in secondary currents in river bends and  confluences. 
 It affects the salt distribution in estuaries, interactions with tidal flows, and at river confluences, the transport of sediments and mixing,  erosion, and morphology \citep{constantinescu_11}. Furthermore, stratification is seen as playing an essential role in the formation and structure of sub-marine turbidity currents, as measured and analyzed in \citet{azpiroz_17}, leading to sediment suspension and transport for long distances. Helicity is produced by  hydrostatic
 and geostrophic balance between the Coriolis force, gravity and pressure gradients, while neglecting the nonlinear advection term \citep{hide_02, marino_13h}. In tropical storms, helicity is associated with the presence of coherent structures in the form of roll vortices in the boundary layer of typhoons, structures which are linked with inflection-point shear instabilities \citep{morrison_05}. Similarly, at river confluences, large-scale turbulence structures  are produced in the form of stream-wise oriented helical vortical eddies. When such structures are observed, the mixing properties downstream of the confluence and the mixing interface between the two volumes of water, sometimes of quite different turbidity and salinity, are altered. Helical structures also interact with shear layers, leading to the formation of turbulent motions \citep{constantinescu_11}. Moreover, helicity can lead to large-scale instabilities \citep{frisch_AKA_87, levina_14b, yokoi_16}, and it has been shown to slow-down the temporal evolution of shear flows as well, necessitating a change in the modeling formalism of the unresolved small scales, by incorporating turbulent transport coefficients that are helicity-dependent  \citep{yokoi_93, baerenzung_08, baerenzung_11}. The role of helicity in sub-grid scale models has also been analyzed recently, numerically  as well as analytically, in the latter case using  upper bounds on weak solutions of the Navier-Stokes equations \citep{linkmann_18}. 

One can define spectral densities of  energy and helicity in terms of isotropic wavenumber $k$,  [$E_V(k), H_V(k)$], and with $\left<{\bf u}\cdot {\bf u}\right>=2\int E_V(k)dk$, $\left<{\bf u}\cdot {\mbox{\boldmath $\omega$}} \right>=2\int H_V(k)dk$ the total kinetic energy and helicity, angular brackets standing for volume integration. With such definitions, one has % a constraint, %namely that 
$ -1\le \sigma_V(k)=H_V(k)/[kE_V(k)]\le 1$
for the so-called relative helicity density $\sigma_V(k)$, using a Cauchy-Schwarz inequality. 
In FDT, both $E_V(k)$ and $H_V(k)$ follow a $k^{-5/3}$ spectral law and thus the return to isotropy in homogeneous isotropic turbulence, with full recovery of mirror symmetry ($\sigma_V(k)\rightarrow 0$), occurs at a rate proportional to $1/k$ as $k\rightarrow \infty$. 
Note that isotropy here is meant solely as an invariance to rotation, but lack of mirror reflections is allowed, as measured by non-zero 
velocity-vorticity correlations.
%helicity. 

The analysis of triad interactions decomposed into two circularly polarized ($\pm$) helical modes as done in \citet{waleffe_92} shows that an inverse transfer of energy to larger scales takes place when the two small-scale modes of a given triad have the same helical polarity. Although overall the energy cascade is to the small scales, as predicted in \citet{kraichnan_73}, this feature was exploited in \citet{biferale_12} by restricting interactions to one-signed helical modes and observing numerically an inverse cascade of energy in that truncated case. This result further led to the  derivation of global regularity in these truncated systems \citep{biferale_13}.
Such a helical decomposition is rather common; it has been used in so-called shell models \citep{lessines_09}, or in the context of the dynamo problem \citep{linkmann_17a}.

A further and rather intriguing result, as analyzed in \citet{alexakis_17}, is that in fact, the total energy flux can be decomposed into three different sub-fluxes that are individually constant in the inertial range, hinting at some, as yet undetermined, invariants and, as pointed out in the paper, leading to additional exact laws in terms of third-order structure functions. Similarly, the helicity flux can be partitioned into two independently constant partial fluxes, indicative that the behavior of such flows is more constrained than thought previously. It is not clear whether this corresponds to the separate invariance of the (two) energies of these modes. Similarly, for MHD turbulence, 
the invariance of total energy and cross helicity, namely $E_T=E_V+E_M$ and $H_C=\left<{\bf u} \cdot {\bf b} \right>$, can be expressed as two definite-positive invariants, in the ideal, non-dissipative case, in terms now of the so-called Els\"asser variables ${\bf z}_\pm={\bf u} \pm {\bf b}$, of energies $E_\pm$. So the question arises as to whether there are also in MHD sub-fluxes that are independently constant in the inertial range, a point open for future research.

Recently, it was also shown in \citet{slomka_17} that, in the context of bacterial suspensions and using the Navier-Stokes equations for the solvent with a stress tensor including higher-order terms modeling the role of the non-Newtonian active part of the fluid, an accumulation of energy at large scales occurred because of an instability due to the bi-Laplacian forcing. One peculiar feature of these solutions is that they can be fully helical, Beltrami flows, ${\mbox{\boldmath $\omega$}} =\lambda {\bf v}$ by selecting the scales for which the three linear terms (proportional in Fourier space to $k^{2n}, \ n=1,2,3$) can balance each other exactly. Note that this corresponds to a viscous-forcing-dissipation balance, somewhat similar to the large-scale quasi-geostrophic equilibrium in rotating stratified flows involving, rather, dispersive effect of inertia-gravity waves and the pressure gradient. Such an inverse energy cascade occurs through spontaneous mirror-symmetry breaking as being the cause of the selection of triadic interactions. Indeed, these new terms in the stress tensor can induce, when the equation is linearized, instabilities of the large scales because of a limited range of unstable Fourier modes depending on the governing parameters.

\section{Coupling to a magnetic field} \label{S:MHD} 
 \subsection{Dynamical equations and parameters}
For completeness, one can write the MHD equations in the incompressible case (see, {\it e.g.}, \citet{pouquet_93, pouquet_96sm, davidson_13, galtier_16}), including the Hall current added to a (generalized) Ohm's law. The Hall MHD (HMHD) equations are:
\begin{eqnarray}
\frac{\partial {\bf v}}{\partial t} &=& -{\bf v} \cdot \nabla {\bf v} - \nabla P + {\bf J} \times {\bf B} + \nu \nabla^2 {\bf v} {- \nu^{\prime}\nabla^4{\bf v}} \label{eq:hall_mom}\\
\frac{\partial {\bf B}}{\partial t} &=& \nabla \times \left( {\bf v} \times {\bf B} \right) - \epsilon_H \nabla \times \left( {\bf J} \times {\bf B} \right) + \eta\nabla^2 {\bf B} \label{eq:hall_ind} {- \eta^{\prime}\nabla^4{\bf B}} \ , \label{eq:hall_b}
\end{eqnarray}
together with $\nabla \cdot {\bf v} = 0 \ , \ \nabla \cdot {\bf B} = 0$.
The magnetic field is expressed in Alfv\'en velocity units, $P$ is the  pressure, and ${\bf J} = \nabla \times {\bf B}$  the current density;
$\nu$ and $\eta$ are the  kinematic viscosity and magnetic diffusivity, and 
 $\nu^{\prime},\ \eta^{\prime}$ are hyperviscosity and hyperdiffusivity coefficients  associated with bi-Laplacian  terms. Their respective roles on the formation of small-scale current sheets are discussed in detail in \citet{stawarz_15H} (see also Figure \ref{f:viz1b}, for which we took 
 $\nu=\eta$, $\nu^{\prime}=0=\eta^{\prime}$.) 

The Reynolds, Froude and Rossby numbers are defined as $Re=U_0L_0/\nu$, $Fr=U_0/[NL_0]$ and $Ro=U_0/[fL_0]$, with $U_0, L_0$ being characteristic velocities and length scales based on the computed {\it rms} velocity and integral length scale. 
They measure the strength of nonlinearities relative to dissipation, stratification or rotation respectively, with $f=2\Omega$, $\Omega$ being the rotation assumed to be in the vertical (z) direction. The buoyancy Reynolds number is ${\cal R}_B=Re Fr^2$.
The magnetic Reynolds number is $R_M=U_0L_0/\eta$ and $P_M=\nu/\eta$ is the magnetic Prandtl number.
The dimensionless parameter $\epsilon_H = d_i/L_0$ giving the ratio of the ion inertial length to the integral length scale of the system characterizes the strength of the Hall current. When $\epsilon_H = 0$, the HMHD equations reduce to the MHD equations, and the
 Navier-Stokes equations are obtained by further setting ${\bf B}=0$. The electron-MHD equations are written separately in \S \ref{ss:electron}. 

 \subsection{The direct cascade of energy to  small scales}
   
 In three-dimensional MHD, the ideal invariants in the absence of dissipation are the total energy 
$E_T=E_V+E_M=1/2 \left<|{\bf u}|^2+|{\bf B}|^2 \right>$, the magnetic helicity 
$H_M=\left<{\bf A} \cdot {\bf B} \right>$ with ${\bf B}=\nabla \times {\bf A}$ (where  ${\bf A}$ is the magnetic vector potential), 
and the cross-correlation between the velocity and magnetic field, $H_C=\left< {\bf u} \cdot {\bf B} \right>$ \citep{woltjer_60, pouquet_96sm, blackman_15}. 
The total energy cascades to small scales, both in two and in three dimensions, with a  self-similar spectrum whose spectral index might still be in dispute \citep{mininni_07b,  lee_10, 
beresnyak_14, perez_14}, a matter that is rendered difficult (i)  by the anisotropy of such flows especially in the small scales because of the presence of a uniform (or quasi-uniform) strong magnetic field in the large scales; (ii)  by the presence of non-zero correlations between the velocity and the magnetic field, as well as (iii) by intermittency effects leading to the steepening of these spectra. Indeed, the intermittency in MHD is known to be stronger than for  FDT, and to be variable as well. For example, the anomalous exponents for the scaling of structure functions depend on the intensity of solar flares \citep{abramenko_10}. Energy spectra have been observed in the Solar Wind consistently over the years (see {\it e.g.} \citet{matthaeus_82, marino_08, veltri_09}). They correspond to a direct energy cascade which leads to plasma heating \citep{stawarz_09, marino_11}, potentially through reconnection events of current and vorticity sheets, as recently detected for example in the Earth's magnetosheath \citep{phan_18}. 
% Since dissipation hasn't been directly linked to reconnection in the solar wind. Also the Phan et al paper finds evidence of reconnection linked with turbulence driven current sheets in the magnetosheath, and while we have observed reconnection in the solar wind, I think it is typically linked with current sheets generated by the large scale structure of the solar wind, as opposed to those linked to the turbulence.  
However, there are a variety of possible mechanisms for the  dissipation of turbulence in collisionless space plasmas, such as  current driven instabilities \citep{stawarz_15j}, and it is still an open question as to what  the dominant mechanisms are. Reconnection is also present in numerous numerical studies (see {\it e.g.} \citet{ting_86, matt_rev_04, mininni_06b, higashimori_13, karimabadi_13, loureiro_13}), and in plasma relaxation processes \citep{taylor_86}. 
 Finally, there are recent indications, using small-scale laboratory experiments, that the source of the turbulence may not matter in the resulting spectral dynamics of such MHD flows \citep{chatterjee_17}.

The observation of magnetic fields in planets and stars, in the interstellar medium or in galaxies leads to the question of their origin. 
Many review papers and books have been devoted to this topic, and one can consult for example \citet{pouquet_93, pouquet_96sm, branden_rev, galtier_16}.
They may be primordial remnants from the early universe, or they may arise from a generation mechanism from a seed field  ${\bf b}$ coupled to the velocity, through what is called the dynamo effect. Reversals of the geomagnetic field occur in a rather reproducible way, on average, with fast growth and slower decay phases that are comparable between different geological periods  \citep{valet_16}. It has been known, since Parker's model and the mean field theory developed in \citet{steenbeck_66}, that helicity is an essential ingredient of such a dynamo mechanism for creating large-scale fields.
 Note however that one can find  examples of non-helical dynamos that can take place through chaotic stretching of magnetic field lines (see {\it e.g.} \citet{childress_95, ponty_95, nore_97}). Once a strong magnetic field is generated, it  leads to wave propagation and  to a  weak turbulence regime that has been detected in the magnetosphere of Jupiter \citep{saur_02}. It also leads to the development of localized vorticity and current sheets which further roll-up as the turbulence increases  for sufficiently high Reynolds numbers \citep{lee_10}, as observed for example in the Earth's magnetosphere \citep{alexandrova_06, osman_14}. 

The decay of energy is lessened in the presence of cross-helicity, {\it i.e.} the correlations between the velocity and the magnetic field \citep{pouquet_88, smith_09, marino_12}. This corresponds again to an alignment, here of the velocity and the magnetic field, which weakens the nonlinear terms, directly for Ohm's law, and indirectly for the Lamb vector ${\bf u}\times {\mbox{\boldmath $\omega$}}$ 
which can be compensated exactly, when the correlations are strong, by the Lorentz force ${\bf B} \times {\bf j}$. 
The role of symmetries in the long-time evolution of MHD flows, using an energy minimization principle, has been shown to lead to final states governed by the relative magnitude of ideal invariants \citep{stribling_90, stribling_91, stawarz_12}. Furthermore, 
 the role of anisotropy is central in the dynamics of the cross-correlation, as shown in \citet{briard_18} using a two-point closure of MHD turbulence, and it should be investigated further. It is known that the return to isotropy in the small scales can be slow, as for example in the presence of rotation \citep{mininni_12}. Recent laboratory experiments \citep{baker_18} and high-resolution numerical simulations on grids of up to 16384x2048$^2$ points % in non-cubic domains 
\citep{zhai_18}, have studied MHD anisotropy and its relationship with dual cascades. It is found that 2D and 3D coherent structures, in the bulk and at the wall, can cohabit. In fact, such a transition also exist in variations of the Navier-Stokes equations having the same inviscid invariants (energy and helicity) and the same symmetries (under rotation, reflection and scaling). It can be analyzed in terms of critical point behavior with divergence of the fluctuations field at the transition \citep{sahoo_17}.

 \subsection{Inverse cascade of magnetic helicity}
 
  %%%%%%%%%%%%%%%%%%%%%%%%%  F1  MHD %%%%%%%%%%%%%%%%%%%%%%%%
 \begin{figure*}    
\includegraphics[width=.47\textwidth, height=12.3pc]{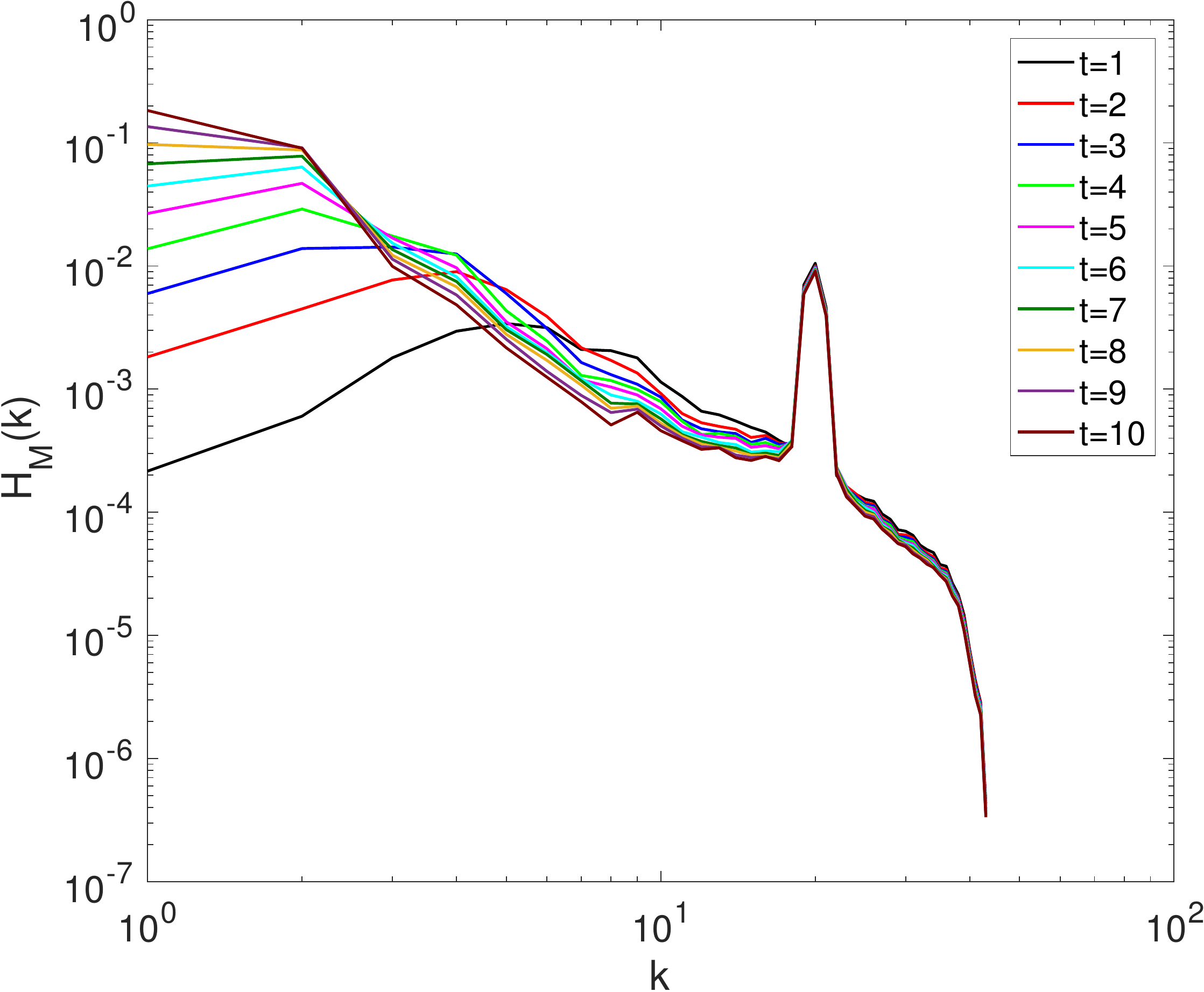}
\includegraphics[width=.47\textwidth, height=12.3pc]{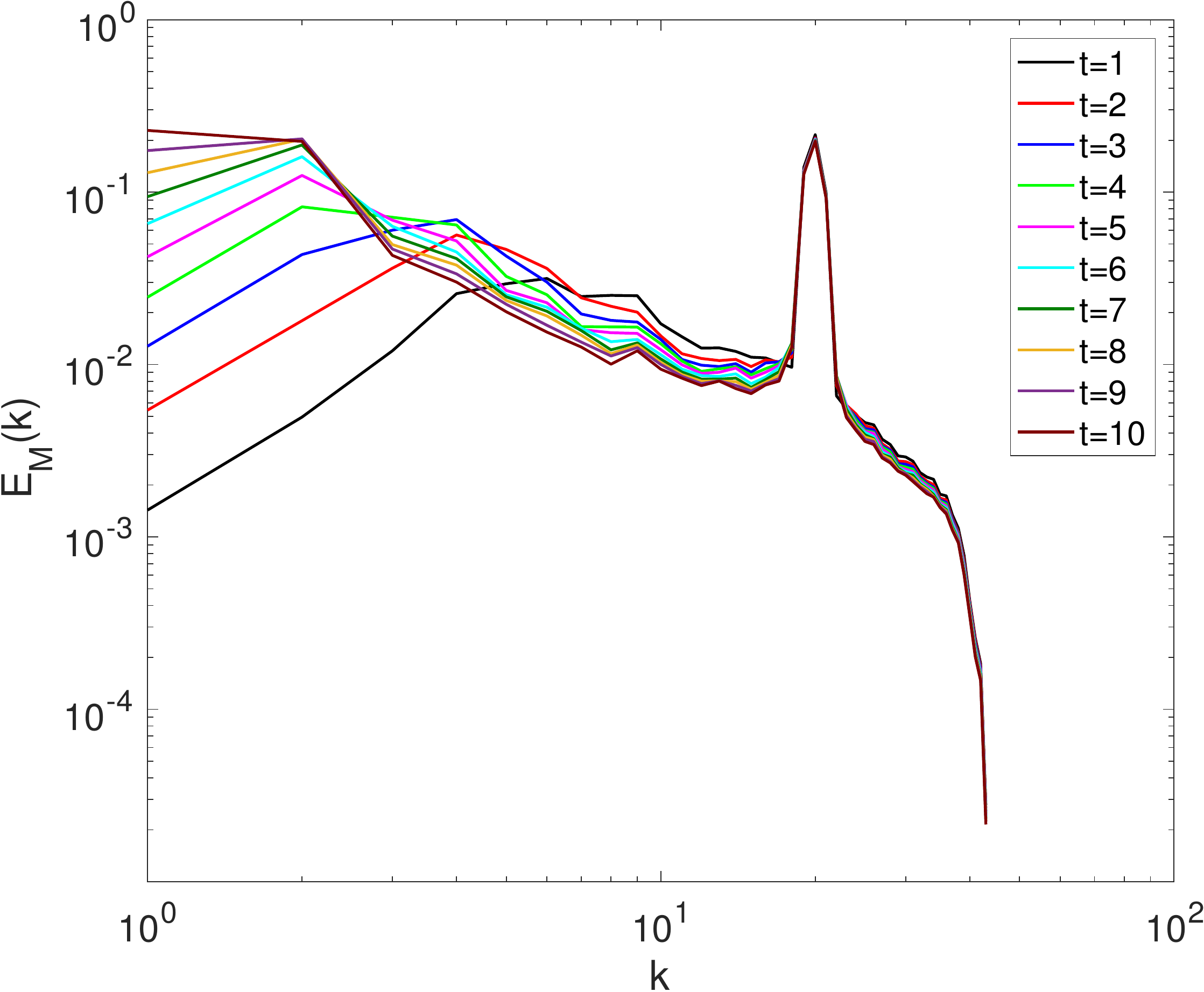}
\includegraphics[width=.47\textwidth, height=12.3pc]{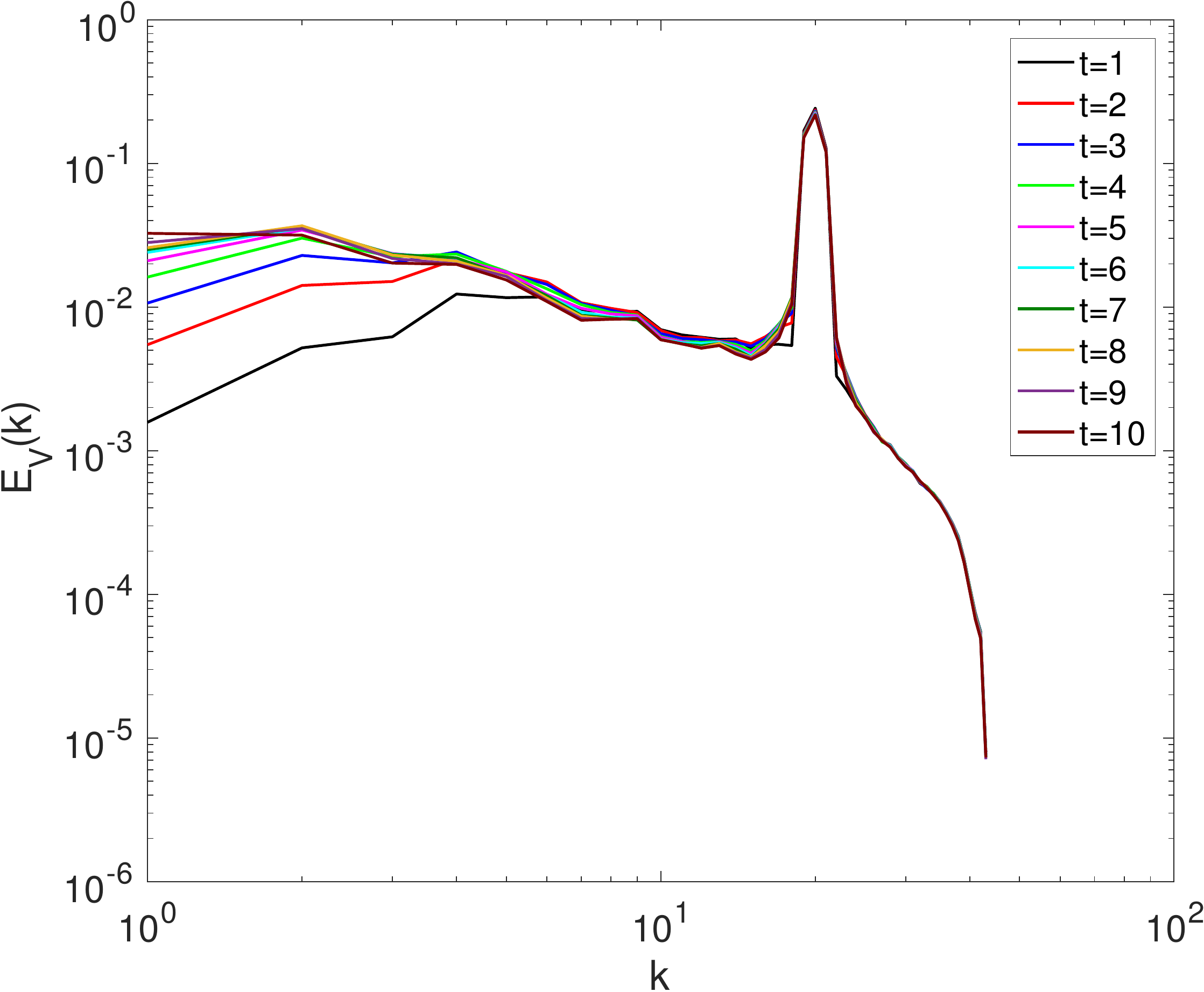}     
\includegraphics[width=.47\textwidth, height=12.7pc]{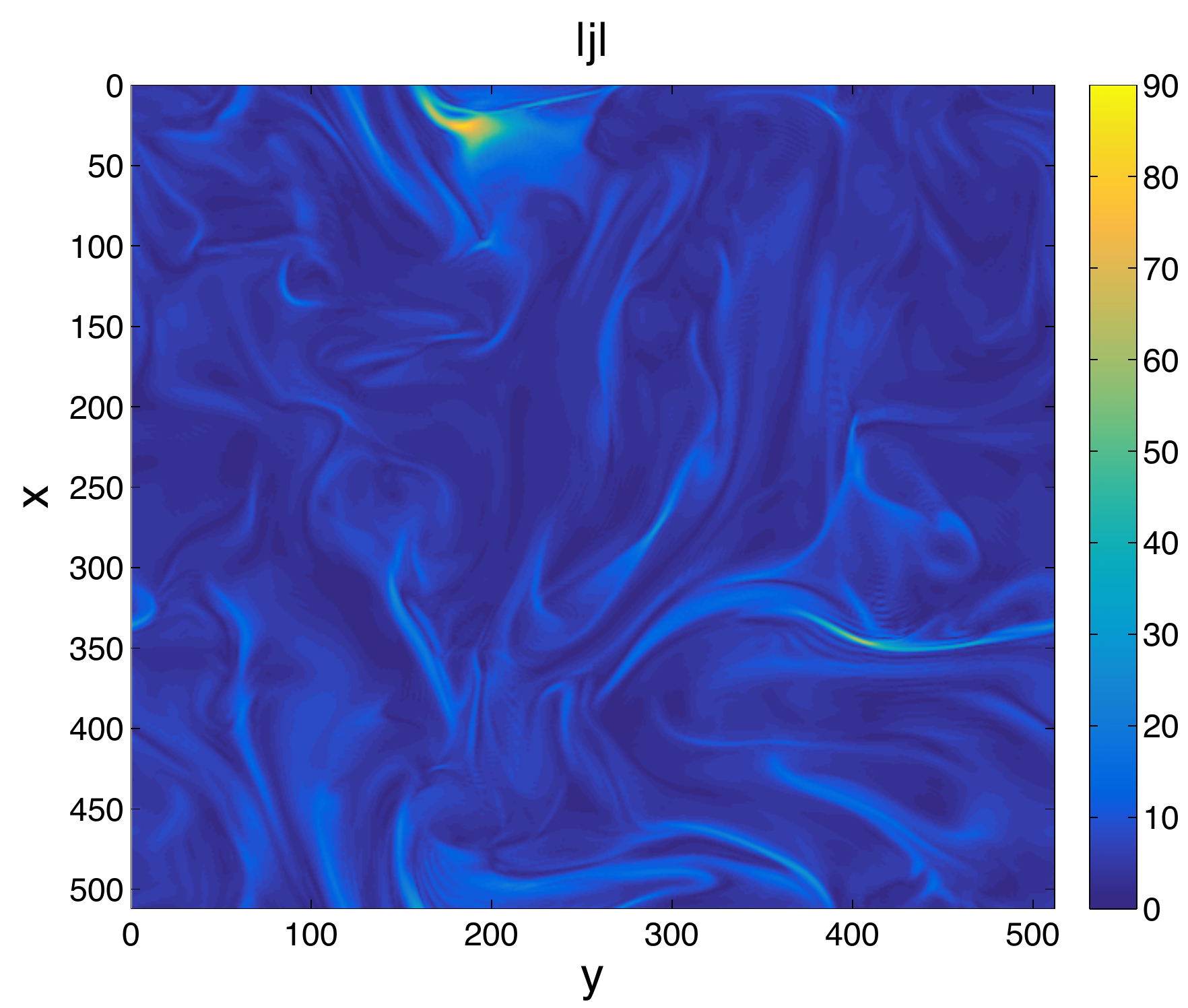}
\caption{
Spectra, drawn at different times (see inset), in forced MHD turbulence for magnetic helicity and energy  (top left and right), and for  kinetic energy (bottom left); new runs on grids of $128^3$ points without hyperviscosity ($\nu^\prime=\eta^\prime=0$ 
in equs. (\ref{eq:hall_mom}, \ref{eq:hall_b})). Note the accumulation of excitation at large scale as time increases.
{\underline{{\it Bottom right:}}} horizontal cut for Hall-MHD of the current density for a new run on a grid of $512^3$ points, again without hyper-diffusivities (see \citet{stawarz_15H} for more detail of the general set-up).}  \label{f:viz1b} \end{figure*}
% , since it looks like those plots were from June of 2015 and I didn't do anything with hyperviscosity until after we received the reviews on the HMHD paper at the end of September 2015. The Hall-MHD plot should also be using normal viscosity since it is $512^3$. The hyperviscous runs were only done up to $256^3$ based on what is the HMHD paper and what I have on my computer.
%%%%%%%%%%%%%%%%%%%%%%%%%%%%%%%%%%%%%%%%%%%%%%%%%%%%%%

Magnetic helicity is observed in  the solar photosphere (see {\it e.g.} \citet{blackman_15} 
and references therein),  and in the Solar Wind \citep{howes_10}. Its dynamical role in coronal mass ejections can be modeled  through the twisting and reconnection of magnetic flux tubes \citep{gibson_08, 
malapaka_13}, and it is known to undergo an inverse cascade  to large scales \citep{frisch_75, pouquet_76, pouquet_93, pouquet_96sm}. 

In Figure \ref{f:viz1b}, we  give the temporal development in MHD of the magnetic energy and helicity spectra, 
$H_M(k), E_M(k)$ (top left and right), and the kinetic energy spectra, $E_V(k)$, bottom left. Forcing is for $k_F\approx 20$; times are given in the inset in units of the turnover time $\tau_{NL}=L_0/U_0$.
The Geophysical High-Order Suite for Turbulence (GHOST) code is used \citep{hybrid_11} with grids of $128^3$ points; boundary conditions are periodic, and only a normal Laplacian operator is present. % for the two diffusive terms.
One observes the progressive build-up of $H_M$ at large scale which entrains the magnetic energy, since $E_M(k)\ge k|H_M(k)|$; this build-up of $E_M(k)$ then leads to a similar growth of kinetic energy at large scales because of the effect of Alfv\'en waves \citep{pouquet_76}.  
The figure here is a simple illustration of the phenomenon. Detailed triadic interactions of such an inverse cascade are studied in \citet{linkmann_17a, linkmann_17p} using a classical helical decomposition, with both analytical and numerical tools at resolutions of 512$^3$ points.
These authors show in particular that the relative signs of kinetic and magnetic helicity play a direct role in the emergence of large-scale dynamo fields, with growth-rates that are determined in the ideal (non-dissipative) case and thus independent of the magnitude of the magnetic Reynolds number, with no subsequent so-called $\alpha$-quenching (see also \citep{pouquet_76}).  

We  show, in the bottom right of Fig. \ref{f:viz1b}, small-scale current structures drawn at the peak of dissipation in the presence of a  Hall current. Here, the run is unforced, with random initial conditions, again no hyperviscosity, and it is performed on a 
more highly-resolved grid of $512^3$ points. Note the strong localized structures.
% An inverse transfer of magnetic helicity is also observed in Hall-MHD \citep{mininni_07}.
 This is discussed further below in the context of 
electron MHD, which is a simplification of HMHD where one assumes that protons are stationary, an hypothesis which is potentially valid for the small scales of Hall MHD.

\subsection{Phenomenology of the inverse cascade for electron MHD} \label{ss:electron}

% One can argue that, since small scales have shorter timescales,  the invariant with the smallest number of derivatives, here the magnetic helicity, goes to the large scales and the other invariant, here the total energy, to the small scales. Such l
Large-scale helical structures in MHD are observed, including in the case when the forcing is non-helical, and their life-time (that is, the time spent on these states) is directly linked to the temporal correlation time of the forcing function \citep{dallas_15}. Because such large-scale energetic structures are force-free, that is the magnetic field and current are parallel and hence the Lorentz force is zero, it poses the question of a possible linearization of turbulent flows around such stable states  in plasma relaxation processes \citep{taylor_86}. 
Magnetic helicity is also detected in the laboratory in perhaps the simplest  instance of the family of models for  plasmas, namely electron MHD (or EMHD) \citep{stenzel_95}. In that case, the velocity of the electrons is slaved to the electric current (while ignoring protons), and the resulting equations only involve the magnetic induction ${\bf B}$; the equations are: %with ${\bf J}=\nabla \times {\bf B}$ the current density:
\begin{equation}
\frac{\partial {\bf B}}{\partial t}  + \alpha \nabla \times [{\bf J} \times {\bf B}] = \eta \nabla^2 {\bf B} \  \ , \ \ \alpha = \frac{c}{4\pi n_e e} \ ,
\label{emhd} \end{equation}
with $c$ the speed of light and $n_e$ the electron density of charge $e$. %, and $\lambda$ the magnetic diffusivity.
In the absence of dissipation, the magnetic energy $E_M$ and the magnetic helicity $H_M$ are conserved, and one has $|\sigma_M(k)|\le 1$, with $\sigma_M(k)=kH_M(k)/E_M(k)$; $\sigma_M=\pm1$ corresponds to the alignment (or anti-alignment) of the vector potential and the magnetic induction. 

As in MHD, waves can travel along an imposed strong magnetic field, in either direction; the so-called {\it unbalanced} case is when more waves travel in one direction than in the other one. Since the waves in EMHD are circularly polarized, the unbalanced case results in a non-zero magnetic helicity, and thus in an inverse cascade of $H_M$ and direct cascade of $E_M$. Using direct numerical simulations 
of decaying EMHD in 3D,  it is shown in \citet{cho_11} that the peak of the magnetic energy spectrum moves to larger scales; this is consistent with the fact that $|\sigma_M(k)|\le 1$, and the magnetic helicity may undergo an inverse transfer as well in the forced case (see also \citet{zhuJZ_14}). However, note that the sign of the energy flux was %indeed 
predicted analytically to be positive in the framework of weak turbulence for EMHD, 
corresponding to a direct energy cascade under the assumption that there is an infra-red (large-scale) cut-off to insure locality of Fourier interactions \citep{galtier_03b} (see also \citet{galtier_15}).

An argument for the inverse cascade can be made following what is done in \citet{fjortoft_53} for two-dimensional neutral fluids, since magnetic energy and helicity are dimensionally linked, with $|\sigma_M(k)|\le 1$. In order to be able to straightforwardly extend the result of Fj\"ortoft in that case, one needs to assume that helicity is of a given sign and that it is maximal (see {\it e.g.} \citet{waleffe_92, chen_03, slomka_17} for detailed analyses of triadic interactions in helical flows). 
This  applies, for example, to the set-up of the inverse energy cascade found for one-sign helical flows \citep{biferale_12}. Moreover, within the large-scale helical range, the helicity is likely to be of one sign only, and in MHD at least it is known to be maximal in the largest scales, corresponding to force-free field configurations \citep{pouquet_76}. It can also be noted that a mechanism for having one-sign magnetic helicity in EMHD at large scale was discussed in \citet{zhuJZ_14}: the argument consists in remarking that, in terms of helical variables, the statistical equilibria are large-scale dominated close to their Fourier-space pole. 
% ``due to the small-k pole of one of the two chiral sectors''

For the EMHD system, let us assume that we have an exchange of magnetic energy and of magnetic helicity for the triplet of wavenumbers $[k,p=1.5k, q=2.25k]$ (which, for integer wavenumbers, can be the triad
[k = 4, p = 6, q = 9]); this exchange is written as $\delta E_{M,kpq}$ and $\delta H_{M,kpq}$, and say that helicity is one-signed, positive,  and maximal, $H_M(k)=E_M(k)/k$. The double constraint of conservation of total $E_M$ and $H_M$ leads to algebraic relations for the energy and helicity interactions between these three wavenumbers. Thus, one obtains:
\begin{equation}
 \delta H_{M,k} + \delta H_{M,p} + \delta H_{M,q} = 0 \ , \label{inv-emhd1} \end{equation}
\begin{equation}
\delta E_{M,k} + \delta E_{M,p} + \delta E_{M,q} = 0 \  =\  k\delta H_{M,k} + p \delta H_{M,p} + q\delta H_{M,q} \ ,
\label{inv-emhd2} \end{equation}
due to the detailed conservation properties within each triadic interactions. In the specific case chosen here, this gives $\delta H_{M,k} =\frac{3}{2} \delta H_{M,q}  \ , \ \delta E_{M,k} =\frac{2}{3} \delta E_{M,q}$.
So we conclude that, under these hypotheses, and starting from an excitation at the intermediate scale $\sim 2\pi/p$, there is more magnetic {\sl helicity} transfer to the large scales $\sim 2\pi/k$, and more magnetic {\sl energy} transfer to the small scales $\sim 2\pi/q$. 
This  agrees with the fact that, in the forced case, the flux of magnetic helicity is observed to build-up an inverse cascade over time \citep{kim_15}. It should be noted that, when injecting energy of $\pm$ circularly polarized EMHD waves in a system through a forcing mechanism at some scale, one necessarily injects magnetic helicity, which can thus be regarded as a direct consequence of the properties of such waves \citep{cho_11}. The ratio of the $\pm$ injection (and dissipation) rates, $R_{emhd}=\epsilon_+/\epsilon_-=D_tE_+/D_tE_-$ appears to be the central parameter determining the fate of these flows: the inverse energy cascade is found for $R_{emhd}\approx 1.2$, but instead of a pure self-similar spectrum for the energy, it takes the form of an envelope \citep{kim_15}. 

The direction of cascades in the presence of more than one invariant is also reviewed and analyzed at length in \citet{alexakis_18}.
 A rather novel result in MHD is that  in fact the magnetic helicity undergoes constant-flux cascades to both large scales and small scales  \citep{alexakis_06, mueller_13}. This was shown using a detailed analysis of the degree of non-locality of nonlinear interactions (see also \citet{debliquy_05}). The flux of magnetic helicity is observed, somewhat remarkably, to remain of a constant sign across the forcing scale, due to the compensating effects of the change of sign of $H_M$ and that of the flux of $H_M$ across that scale. The nonlocal interactions in MHD are therefore able to smooth-out the process of transfer across the scales. Using Particle In Cell numerical simulations for two-dimensional plasmas including all three components of the fields, it is argued in \citet{che_14}, in the context of Solar Wind observations \citep{marino_08,  marino_11}, that a dual magnetic energy cascade is observed which  is interpreted as being due to wave-wave interactions,  kinetic Alfv\'en waves and whistler waves,  which feed both the electron and ion scales due to the anisotropy of the electric and magnetic fields, although no energy fluxes are given to confirm this finding. Anisotropy plays an essential role in this mechanism since the momentum transfer is shown to occur between the perpendicular and parallel components of the magnetic field, each transferring to either larger or smaller scales. In the latter case, this occurs together with the formation of kinetic-scale micro-current structures, as recently observed \citet{ergun_18, phan_18}. 
% in terms of reference associated with kinetic scale current structure in turbulent plasmas you could maybe reference Phan et al. 2018 (which is on reconnection at kinetic scale current stuctures in the turbulent magnetosheath) and Ergun et al. 2018 (which looks kinetic scale electric fields and currents in the turbulent magnetotail). Both of those are already cited in the paper. 

\subsection{The two-dimensional case in MHD}

In two dimensions,  the purely magnetic invariant, $\left< A^2 \right>$, is positive definite. 
In that case, it can be demonstrated phenomenologically, and numerically as well, that the transition between a fluid-dominated to a magnetically-dominated regime in the presence of forcing is controlled by the ratio of the kinetic to magnetic energy injection rate, namely $\mu=\epsilon_M/\epsilon_V=D_tE_M/D_tE_V$  \citep{
seshasanayan_16}. The two-dimensionality of the flow allows for large numerical resolutions, and the critical states are identified explicitly, with the divergence of a susceptibility at the critical ``temperatures'' (here, $\mu$), and power-law scaling close to the critical points. These points are slightly different for the change of direct to inverse cascade of magnetic potential and that of kinetic energy. 

Specifically, for small magnetic forcing, in this 2D case, there is an inverse cascade of kinetic energy, and the large-scale magnetic energy spectrum corresponds to equipartition of magnetic potential modes (or $E_M(k)\sim k^{+3}$). On the other hand, for strong magnetic forcing, the equipartition is observed in the large-scale kinetic energy, whereas the magnetic potential undergoes an inverse cascade,  as first derived in \citet{fyfe_76} and found using two-point closure techniques for turbulence (see review in \citet{pouquet_96sm}). 
These differences in spectral behavior have their counterpart in configuration space, with a change in dominant structures according to the value of the critical parameter $\mu$, and in the amount of intermittency  as measured through the exact cubic flux relationships arising from the conservation laws of energy for the two different underlying systems  \citep{seshasanayan_16} 
(see \citet{politano_98b, gomez_00, politano_03} for the 3D MHD helical case). 
It would be of interest to compute as well high-order moments corresponding to the fat tails in the Probability Distribution Functions that are observed to quantify the multi-fractality of these systems, and its disappearance in the inverse cascade. One of the conclusions  in \citet{seshasanayan_16} is that there is a range of values of $\mu$ for which two regimes cohabit, one mostly fluid, one mostly MHD and with at the same time a flux of energy which is both positive and constant, corresponding to the direct cascade, and negative and constant corresponding to the inverse cascade. Furthermore, two-scale instabilities in both 3D  and 2D  using Floquet analysis are identified in \citet{alexakis_18b}. A stability diagram is built and an overlap region of the two types of instabilities in the laminar case is found; it can be linked to the turbulent bi-directional cascade phenomenon as a function of the height of the fluid.

The  two-dimensional case in the absence of a third component of ${\bf u}$ and ${\bf B}$ is a special subset of the more general case,
in the sense that it does not allow for helical (topological) invariants, since the flow is horizontal (say) and the resulting vorticity is vertical.
 However, when including the third component of the fields (vertical velocity and magnetic field), the helicity is non-zero {\it a priori}, more invariants arise \citep{montgomery_82} and this would deserve further study since this configuration also corresponds to three-dimensional MHD in the presence of a strong uniform magnetic field (see {\it e.g.}, \citet{linkmann_18b}). Also note that the spectral index of the inverse cascade can depend on the anisotropy of the forcing for rotating stratified turbulence \citep{oks_17}.

Finally, for a 2D neutral fluid, one can show that, in the presence of a mean flow (such as vortices or jets) and turbulent fluctuations, the mean momentum stress is proportional to the mean shear  \citep{frishman_17b}.  % PoF NOT PRF
% and that the turbulent energy at long times is directly related to the advection of the mean flow
The mean flow energy is obtained from a balance between the large-scale friction and the turbulent dissipation at small scale \citep{frishman_18}.

\subsection{The role of anisotropy}

There are several issues that merit further attention. One of them   concerns the development of anisotropy in such flows. An imposed uniform rotation or  uniform magnetic field, or gravity, all render the flow quasi-bi-dimensional.  
For rotating or for stratified flows, isotropy is recovered at small scale, beyond what is called the Ozmidov or Zeman scale, 
$\ell_{Oz}=[\epsilon_V/N^3]^{1/2}, \ell_{Ze}=[\epsilon_V/f^3]^{1/2}$, with $N, f=2\Omega$ the Brunt-V\"ais\"al\"a frequency and twice the rotation frequency. At these scales, the characteristic times of a wave and of a turbulent eddy are comparable, with the assumption of an isotropic Kolmogorov (1941) spectrum being recovered at small scale, and that at smaller scales, nonlinear eddies are faster. On the other hand, in the case of MHD, the opposite happens and anisotropy develops as small-scales become more two-dimensional. A uniform magnetic field $B_0$ has a strong influence on small-scale dynamics, whereas a uniform velocity can be eliminated through Galilean invariance. This represents a difference between purely fluid and MHD turbulence.
The resulting anisotropic bi-directional cascade has been studied in  detail, with a progressive increase of the inverse (quasi-2D) flux as $B_0$ increases (see also \citet{favier_11, reddy_14, gallet_15}, and for the two-dimensional case,  \citet{shebalin_83}).
  It helps in  the interpretation of Solar Wind data \citep{verdini_15}, and has also applications in metallurgy: 
indeed, it is known that for sufficiently strong $B_0$, the fluid behaves like a quasi 2D three-component neutral fluid, the Lorentz force acting as an anisotropic dissipation \citep{garnier_81}, together with nonlocal energy transfers between the toroidal and poloidal components of the fields \citep{favier_11, reddy_14}. It is further found that the perpendicular components of the velocity undergo an upscale cascade whereas its parallel component follows a direct cascade, although this may depend on the strength of $B_0$. Such phenomena display a critical behavior  \citep{sujovolsky_16}, the intensity of the magnetic fluctuations increasing as the one-half power of the Alfv\'en time based on the large-scale flow and on the uniform field. Furthermore, for strong $B_0$, the direct cascade is dominated by  helicity (in relative terms), as in the rotating case \citep{mininni_09c}. It is not clear what happens when both rotation and an imposed uniform magnetic field are present, but the angle between these two imposed fields may  alter the dynamics in significant ways. For example, it is shown in \citet{salhi_17} that the Alfv\'en ratio (that is, the ratio of kinetic to magnetic energy in terms of radial spectra) has a different power-law decay for the parallel and orthogonal cases. 
It is seen moreover that quasi-equipartition recovers for wavenumbers such that the parameter $V_A k$ be much larger than unity, with $V_A=\sqrt{B_0^2/[\rho_0 \mu_0]}$ the Alfv\'en velocity associated with the imposed magnetic field $B_0$, where $\rho_0$ is the  density taken as uniform, and $\mu_0$ is the permeability of the vacuum.
 On the other hand, at large scale, the equipartition is between the kinetic and potential energy for $V_Ak/N$ much smaller than unity.
%with $N$ Brunt-V\"ais\"al\"a frequency.

 \section{The bi-directional, or dual, cascades in turbulence} % rotating stratified flows}

Dual cascades have been observed in a variety of contexts, as in the ocean \citep{scott_05, sasaki_14, balwada_16, klein_19},
 in observations of the Solar Wind \citep{sorriso_07} and of Jupiter \citep{young_17}, in experiments in a rotating tank \citep{morize_05} as well as in DNS of MHD flows \citep{alexakis_11} or of rotating flows \citep{kafiabad_16}. The combination of direct and inverse energy cascades is invoked in the evolution and regeneration of wall turbulence \citep{farano_17}, with the lifting of coherent structures in large-scale hairpin vortices which further destabilize and renew the (optimal) bursting cycles of such flows.

% \cite{klein_18} ESS interactions between surface dynamics and in depth about the scale at which the inverse cascade of energy takes place, as a need to reinterpret the flux data from ocean floaters.
 
Rotating stratified turbulence (RST) is strongly anisotropic  (see {\it e.g.}, \citet{cambon_89, cambon_99, favier_10}). In such flows, similarly to the MHD case, there is also a bi-directional constant-flux system of energy cascades  to both the small scales and the large scales \citep{pouquet_13p, marino_15p, pouquet_17p}.  One can summarize these results, by analogy to MHD, in the following manner. 
In the atmosphere and the oceans, energy  is injected  through solar radiation, tides, bottom topography or winds for example. The large scales of such flows are in  hydrostatic balance and in geostrophic balance between  pressure gradient, Coriolis force and gravity. This leads to a quasi-2D behavior with the energy flowing to the large scales and thus without a clear way to dissipate the energy, a process which mostly takes place at small scale. What is, then, the internal mechanism for energy dissipation  in such flows,
outside of boundary layers, or considering the classical Ekman drag?
In the presence of rotation, gravity as well as shear, several instabilities are known to exist; they take their energy from either the kinetic or the potential modes (see \citet{wang_14, feraco_18}). These instabilities can be viewed as the prelude to a direct cascade of energy to the small scales, with a constant flux, together with the inverse cascade. For atmospheric and oceanic dynamics, this split  process for the energy cascade  is essential. 
 
\begin{figure*}      \vskip-0.34truein \hskip0.2truein
\includegraphics[width=.97\textwidth, height=32.2pc]{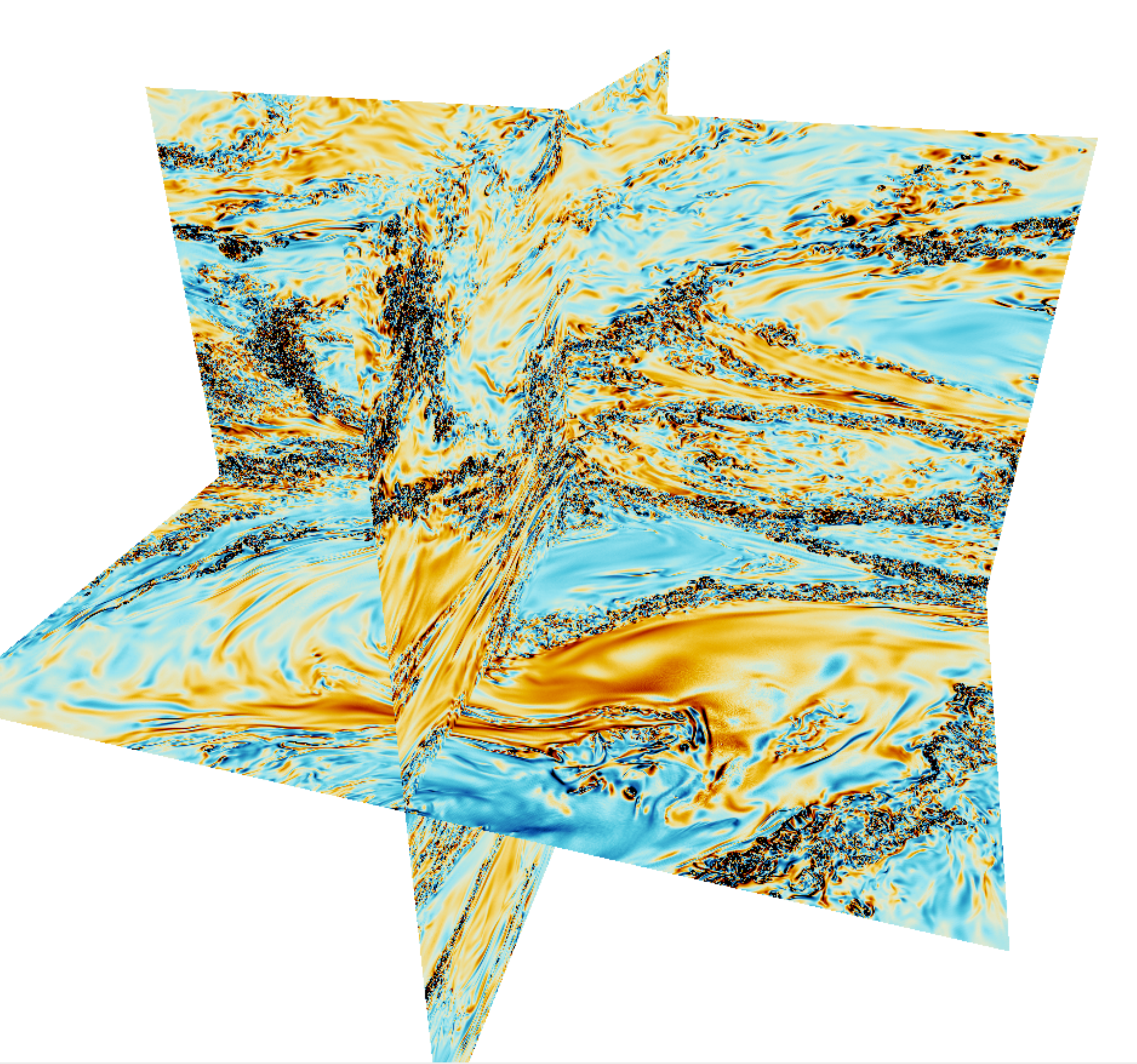}
 \includegraphics[width=.33\textwidth, height=11.9pc]{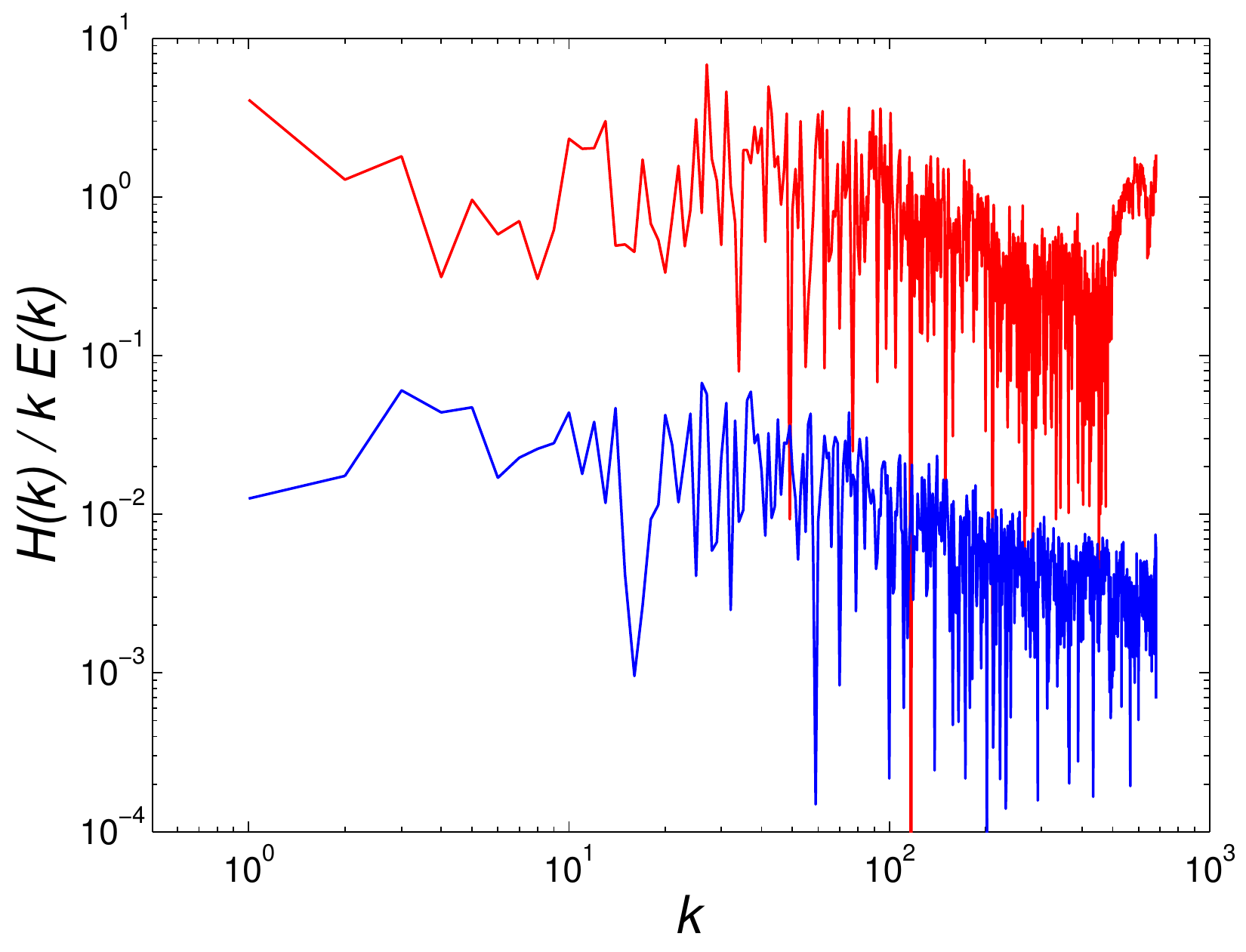}   \hskip0.01truein  
\includegraphics[width=.33\textwidth, height=12.3pc]{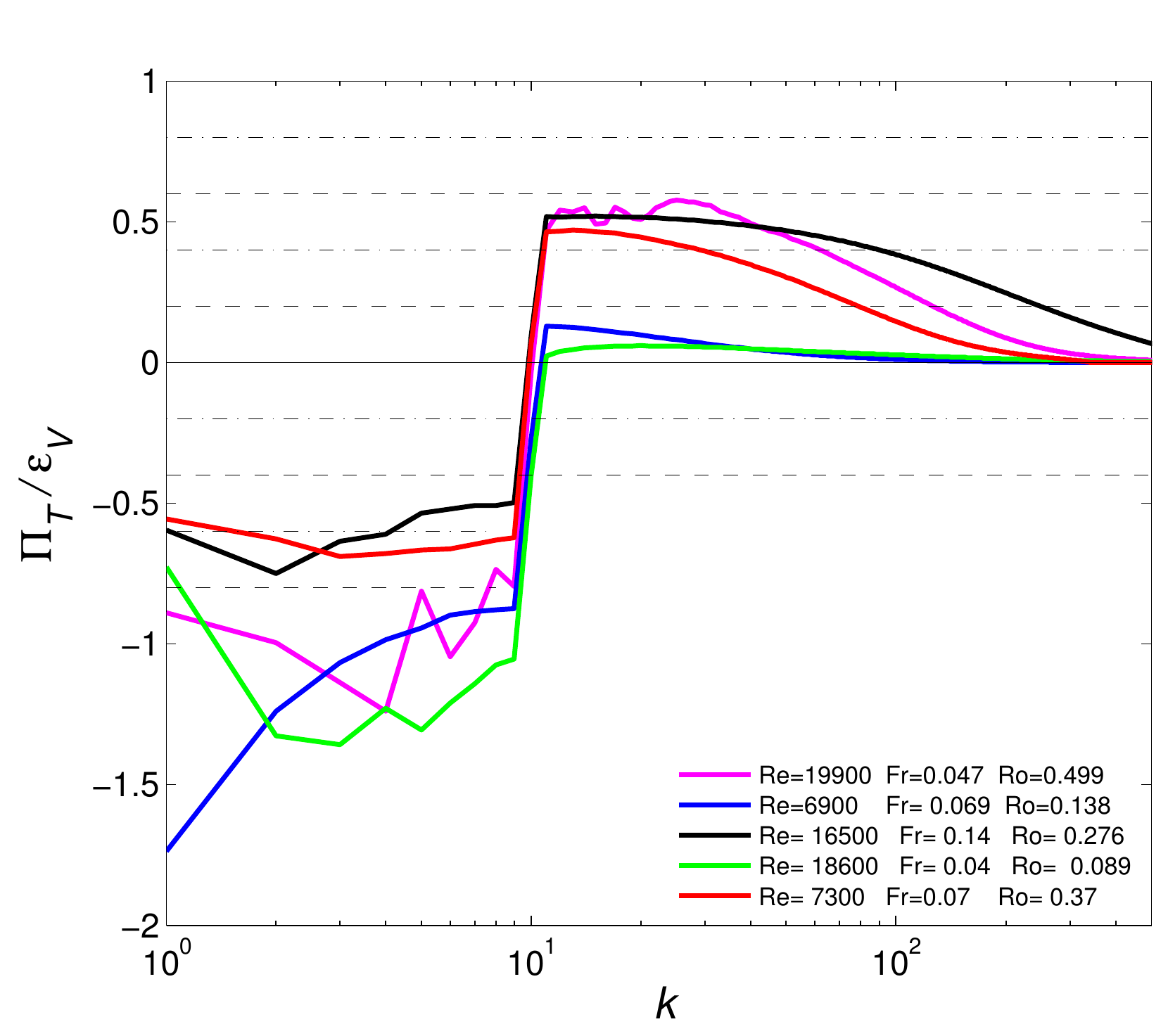}  \hskip0.01truein  
 \includegraphics[width=.33\textwidth, height=12.57pc]{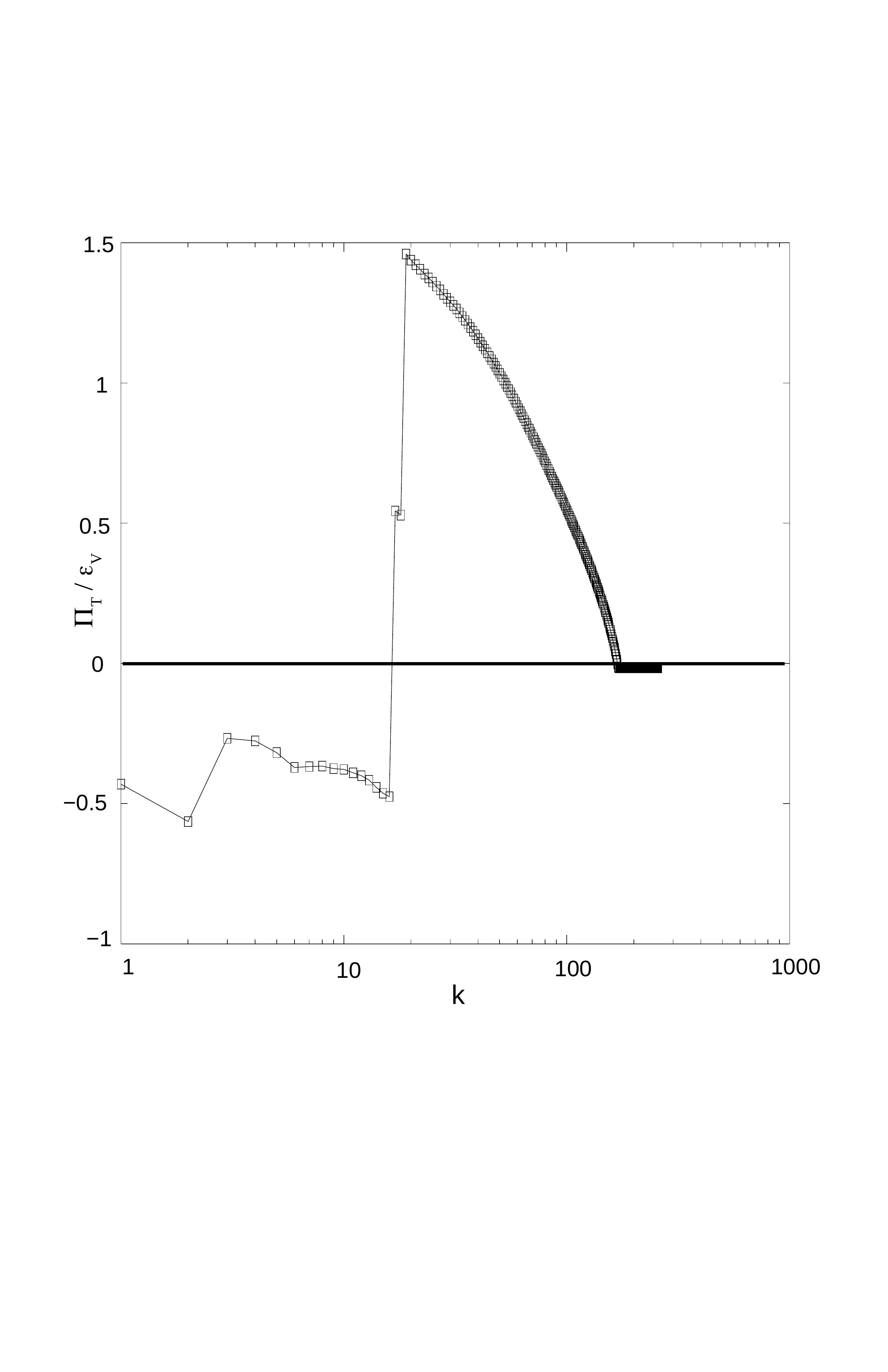}  
 \vskip-0.1truein   \caption{{\underline{Rotating stratified flows}}. 
{\it Top:} Vorticity strength on three planes for the decay run analyzed in \citet{rosenberg_15}, where the  color map is also given. The grid has $4096^3$ points, $N/f=5$, $Re=55,000$, $Fr= 0.024$. 
{\it Bottom left:} Relative helicity spectra at two different times for runs analyzed in \citet{marino_15p} on grids of $2048^3$ points,
$Re \approx 1.65 \times 10^4$, $N/f=2$, $Fr \approx 0.14$. For clarity, the data has been shifted upward by a factor 100 for the earlier time (in red).
{\it Bottom middle:}  Energy fluxes with dual cascades for isotropic forcing in the velocity only, grids of $1024^3$ or $2048^3$ points for runs analyzed in \citet{pouquet_13p, marino_15p}; the parameters are given in the inset.
{\it Bottom right:} Dual cascade energy flux for quasi-geostrophic forcing for a new run;  grid of $512^3$ points, $Re\approx 2500$, $Fr\approx 0.09$, $N/f=1.97$. All fluxes are averaged for several turn-over times after the onset of the inverse cascade.
}  \label{f:viz2} \end{figure*}

In Figure \ref{f:viz2} (top), we plot vorticity structures in rotating stratified turbulence with large-scale eddies due to the influence of  rotation, at the border of which  strong small-scale vortex lanes develop because of instabilities such as Kelvin-Helmoltz or those due to shear. The plot is for a flow with $Re\approx 5.5\times 10^4, Fr\approx 0.024, Ro\approx 0.1, {\cal R}_B\approx 31$, for a decay run on a grid of $4096^3$ points, integrating the Boussinesq equations (see \citet{rosenberg_15} and references therein for more details).

For a run with forcing centered on $k_F\approx 10$ and a resolution of $2048^3$ grid points (see \citet{marino_15p}), the relative helicity 
$\sigma_V(k)$ is displayed in Fig. \ref{f:viz2} (bottom left) at two different times, with the data displaced upward by a factor of 100 
for clarity for the earlier time.
Note that for this run, the buoyancy Reynolds number, ${\cal R}_B=Re Fr^2\approx 313$ is quite high and the small-scale turbulence beyond the Ozmidov scale is close to the isotropic case.
We see that $\sigma_V(k)$ is quasi independent of wavenumber at large scales, in the inverse cascade, whereas it decays as $\approx 1/k$ at scales smaller than the forcing, as it would for fully developed turbulence. 

The bi-directional constant total energy flux, normalized by the kinetic energy dissipation 
$\epsilon_V=\nu \left<|{\mbox{\boldmath $\omega$}}^2| \right>$, is displayed for several runs with parameters given in the inset in Fig. \ref{f:viz2} (middle plot). Buoyancy Reynolds numbers vary between roughly $31$ and $313$ (black curve), with Reynolds numbers high enough that the small-scale fluxes are rather well-developed and approximately constant.
As analyzed in \cite{pouquet_13p, marino_15p, pouquet_17p}, the ratio of the forward to inverse flux varies as $[RoFr]^{-1}$ as long as the Rossby number is below unity and for  buoyancy Reynolds numbers ${\cal R}_B$ above a threshold of order 10. Note that the  analysis of the forced case was also shown to be compatible with the energy flux to the small scales being proportional to $Fr$, when the (total) energy flux to the large scale was inversely proportional to the Rossby number $Ro$: at fixed Froude number, the stronger the rotation, the more the energy flows to the large scales. 

Finally, at the bottom right of Fig. \ref{f:viz2} is given again the total energy flux, but this time with a forcing in quasi-geostrophic balance. 
This run is forced with a classical Taylor-Green (TG) flow with ${\bf F^u}_{TG}= [\sin x \cos y \cos z, -\cos x \sin y \cos z, 0]$.
${\bf F^u}_{TG}$ is incompressible, has zero vertical component and zero total helicity. The forcing in the temperature equation is such that it is in geostrophic balance with the TG flow (see \citet{rosenberg_15}). The dimensionless parameters for this run are $Re\approx 2500$, $Fr\approx 0.09$, ${\cal R}_B\approx 20$, $N/f\approx 2.02$, $Ro\approx  0.18$, with $N=18.6$, and the grid resolution is $512^3$. The run is comparable to the forced run 10e studied in \citet{pouquet_13p, marino_15p}, but at a lower Reynolds number; thus, the direct flux is competing with dissipation and is not constant at that resolution, but it confirms the generality of the bi-directional cascades in RST for different initial conditions.

It is becoming increasingly clear that the role of large-scale shear is central in many turbulent flows \citep{pumir_96}, as in the structure of late-time coherent eddies in 2D \citep{frishman_17b, frishman_18}, in the destabilization of stratified flows as discussed above (see Fig. \ref{f:viz2}, top), or when it leads to the formation of fronts between meta-stable and stable states, or between quiet and turbulent regions \citep{pomeau_86, waleffe_97}. 
The connection, in the context of the dynamics between large-scale predator (the shear flow) and  prey (the turbulent eddies), and in which the details of the small-scale turbulent eddies are rather irrelevant, allows for the classification of turbulent flows by analogy with directed percolation  \citep{pomeau_15, barkley_16}, as supported by laboratory experiments \citep{sano_16}. Indeed, the critical transition to a turbulent state has exponents comparable to those of directed percolation, concerning for example the scaling of the number of active sites in terms of distance to the critical governing parameter of the transition. 
Such an analysis also shows the importance of large-scale zonal flows \citep{shih_16}, of critical layers  \citep{park_18}, as well as  that of regeneration self-sustaining cycles in such flows \citep{waleffe_97, kawahara_01} with, citing \citet{mckeon_17}, ``{\sl the importance of scale interactions in sustaining wall turbulence through the non-linear term}.''

%%%%%%%%%%%%%%%%%%%%%%%%%  F2   RST  %%%%%%%%%%%%%%%%%%%%%%%%
%%%%%%%%%%%%%%%%%%%%%%%%%%%%%%%%%%%%%%%%%%%%%%%%%%%%%%

In terms of oceanic dynamics, the instabilities giving rise to the forward transfer of energy are often identified as sub-meso-scale currents at scales of 1$km$ which have been recently discovered in the ocean  (see \citet{mcwilliams_16} for a recent review), and which, in general, are hard to obtain numerically for lack of resolution in global simulations. They may occur in the boundary layers and, if unstable (which they likely are) will provide a path to small dissipative scales,  taking the form of eddies, fronts or jets, and filaments. 
The oceanic observations analyzed in \citet{arbic_13} indicate as well the existence of a split  bi-directional
flux. Such a dual-cascade system can also be viewed as two independent cascades corresponding to the geostrophic and ageostrophic parts of the flow \citep{bartello_95}. 

In fully developed turbulent flows, it is shown in \citet{ishihara_16, djenidi_17} that the effective amount of kinetic energy dissipation, measured in terms of its dimensional evaluation $\epsilon_D=U_0^3/L_0$, %$\beta$ 
tends to a constant close to $1/2$ for a variety of laboratory and numerical experiments, in the forced case as well as for decaying flows provided their Taylor Reynolds numbers be sufficiently high, but there is also a marked deficiency, of the order of 10\%, in the presence of non-zero helicity \citep{linkmann_18}.
The actual amount of dissipation (as opposed to that expected for a field composed of a superposition of linear waves interacting weakly), directly influences the overall atmospheric and oceanic energetic exchanges, and it can also modify conditions for acoustic transmission, as well as for deep-water drilling. Dissipation and wave effects are thus important features of such flows to be determined. The presence of waves affects directly turbulent flows in lessening the rate at which  energy is dissipated. It has  been observed in several instances \citep{alexakis_13, campagne_16, maffioli_16b, pouquet_17p, feraco_18}, with in the case of rotating stratified flows, a clear linear dependence of the measured dissipation on the control parameter, the Froude number in an intermediate regime of eddy-wave interactions \citep{marino_15p, pouquet_18}.
On the other hand, the Rossby number controls the amount of energy going to the large scales in a simultaneous inverse cascade, 
whereas for MHD the control parameter is the magnitude of the large-scale imposed magnetic field. It is also shown, both for MHD \citep{alexakis_13} and for RST \citep{rosenberg_16, pouquet_18}  that the two forms of dissipation (kinetic and magnetic, or kinetic and potential) can become comparable. %  for sufficiently high 
In convectively-forced rotating turbulence, using experiments with helium performed in a cylinder of aspect ratio 1/2, \citet{ecke_14} find a similar linear scaling  for the flux measured through the Nusselt number (normalized by its non-rotating value for the same parameters).
Finally, we note that the transition between regimes where either vortices or waves are the dominant modes can be analyzed, in the case of purely rotating flows, in terms of the elliptical instability \citep{lereun_17}. This  can justify the presence of turbulence in planetary interiors, which can in turn lead to a dynamo mechanism for the generation of planetary magnetic fields.

\section{Conclusion and Perspectives}

The study of turbulence, for fluids and  MHD, and of nonlinear systems in general, is progressing substantially as shown, among others, in the few examples given in the preceding sections.
Turbulence studies also lead to the existence of large data sets. An  existing approach to reduce the complexity of the problem has been to identify regions of strong turbulent activity in coherent structures, such as with a principal orthogonal decomposition (see  \citet{mezic_13} for a recent review), through intermittency or by examining enhanced dissipation such as in current sheets, and concentrate the analysis on such regions. When contemplating the vorticity field \citep{ishihara_16}, computed on a grid of  4096$^3$ points (or in excess of 64 billion points), it becomes apparent that the information content of such a picture is enormous, and  the dynamics of the fields on this large grid is computed with spectral accuracy. Analysis of such large data sets is cumbersome.
 Local mesh refinement can help in the analysis of sporadic, intermittent structures which are at the basis of the strong and recurrent  increases in kurtosis in stratified flows \citep{feraco_18}. For example, a comparison of spectral element and finite difference methods using statically refined nonconforming grids for the {MHD} island coalescence instability problem led to the conclusion that a high degree of accuracy is, perhaps unsurprisingly, helpful in identifying correctly the spatio-temporal localization of strong current structures  \citep{ng_08}. 

 Can we do better than increasing the grid size? In other words, is there another way out? Little has been done in the field of turbulence using techniques such as data neural networks, although we already have some pieces of information.  Machine Learning (ML) has now been used in fluid turbulence \citep{duriez_17}, with possible applications to drag reduction for cars, trucks and ships  through active turbulent flow control mechanisms. Other applications include the re-``discovery'' of the underlying equations from experimental and/or numerical data \citep{brunton_17}, the development of new and improved sub-grid scale modeling \citep{ling_16, kutz_17},  the prediction of long-time behavior of chaotic systems \citep{pathak_18}, and specifically that of earthquakes \citep{rouet_17, devries_18}.

Various physics-based model enhancements have already been proposed, as for example incorporating helicity-dependent eddy diffusivities  \citep{baerenzung_08} including in the case of shear flows \citep{yokoi_16}, or when helicity leads to large-scale instabilities \citep{frisch_AKA_87}, or for models including the potential energy for RST \citep{zilitinkevich_08}. One application will be to illuminate the possible connections to climate between atmospheric and oceanic dynamics, as well as human interferences % \citep{steffen_15}. 
   \citep{rockstrom_09, steffen_15}. 

Similarly, the controlling of uncertainties in Reynolds-Averaged Navier-Stokes high-fidelity models can be accomplished through the use of large numerical simulations, either directly exploiting  large turbulence data bases \citep{graham_16, duraisamy_18, wang_17}, or with Large-Eddy Simulations. It may also inform the functional form of the closure schemes, as well as the actual values of the coefficients appearing in such closures. Extreme weather events can also be predicted using instantaneous dynamical system metrics, linked in particular to the rapidity of nearby trajectories in phase-space to diverge \citep{messori_17}.

These  approaches are rather unexplored in MHD.   Automatized current structure identification was performed in \citet{servidio_10, klimas_17} for 2D MHD flows by detecting nulls of the magnetic field with a local hyperbolic topology as the plausible locus of reconnection events. Similarly, one can observe geo-dynamo field reversals using a decomposition of the data through a  chaotic forcing with strong intermittent bursts \citep{brunton_17}. Finally, Lagrangian models (also called $\alpha$-models) allow for computations at high Reynolds numbers by introducing a filter length-scale. These models can, for example, lead to accurate evaluations of sign-cancellations in 2D MHD, leading to the prediction of that exponent being independent of Reynolds number at high Reynolds numbers  \citep{graham_05}. However, this sharp truncation using the $\alpha$ parameter 
may lead to extended regions of very weak nonlinear transfer \citep{mininni_05p}, and Machine Learning procedures could alleviate that problem. One could also think of  incorporating cross-helicity and/or magnetic helicity in transport coefficients for MHD, following similar venues.

Artificial intelligence has been used recently in helping develop more efficient fusion devices that avoid disrupting the magnetically-confined plasmas (see \citet{camporeale_18} for space weather). Data-driven predictions of large-scale disruptions, now at a level of 90\% accuracy,  does allow for early plasma-device shut-down through pattern recognition applied to previous events (there is close to half a peta-byte of data). These disruptions are associated with wall-driven MHD resistive instabilities \citep{vega_14}, and such a classification of  events can be physics-based and multi-dimensional \citep{tang_17}. The disruptions are also related to the burstiness of the turbulence leading to localized anomalous transport in plasmas \citep{diamond_94}, which can be modeled through avalanche dynamics \citep{hahm_18}. 

In what way will existing machine learning algorithms change in the presence of inverse cascades? How will such modeling be affected by the existence of bi-directional cascades? Much remains to be learned but it seems rather certain that Big  Data and Data Science will play a role in MHD turbulence in perhaps much the same way as it has demonstrated its utility in other fields.
These issues could be investigated soon with the existence, as mentioned earlier, of new observational small-scale data with the Mulstiscale Magnetospheric mission \citep{wilder_16, chasapis_17, ergun_18, gershman_18}), with experimental data ({\it e.g.}, coming from the Coriolis table \citep{aubourg_17}), and with numerical data stemming from high-performance computing studies of FDT \citep{ishihara_16}, of rotating and/or stratified turbulence \citep{debruynkops_15, rosenberg_15}, as well as of turbulent interfaces in RST \citep{watanabe_16}, and of MHD turbulence \citep{lee_10, beresnyak_14, zhai_18}.

\begin{acknowledgments}
\noindent
Support for AP from LASP, and in particular from Bob Ergun, is gratefully acknowledged. 
JES is supported by STFC(UK) grant ST/N000692/1. 
RM acknowledges financial support from PALSE (Programme Avenir Lyon Saint-Etienne) of the University of Lyon, in the framework of the program ``Investissements d'Avenir" (ANR-11-IDEX-0007).  
Computations were performed at LASP, on the Janus supercomputer at the University of Colorado (Boulder) for Fig. 1, at DOE for Fig. 2 (top), and on local servers for Fig. 2 (bottom). We thank all such centers. NCAR is supported by NSF. 
 We want to remark that a comprehensive review on closely related topics has just appeared \citep{alexakis_18}. Finally, data 
 for Fig. 2 (top) is available on the Turbulence data base at John Hopkins University \citep{graham_16}; the much smaller data sets used for the other plots are available directly from the authors.
\end{acknowledgments}  

 \bibliography{ap_19_feb19} 
  \end{document}